\documentclass{article}
\usepackage{spconf,amsmath,graphicx,makecell}
\usepackage{balance}

\title{Personalized Audio Quality Preference Prediction}
\name{Chung-Che Wang, Yu-Chun Lin, Yu-Teng Hsu, Jyh-Shing Roger Jang}
\address{
  Dept. of Computer Science and Information Engineering, National Taiwan Univ., Taiwan}

\begin{document}
%
\maketitle
\begin{abstract}
This paper proposes to use both audio input and subject information to predict the personalized preference of two audio segments with the same content in different qualities. A siamese network is used to compare the inputs and predict the preference. Several different structures for each side of the siamese network are investigated, and an LDNet with PANNs' CNN6 as the encoder and a multi-layer perceptron block as the decoder outperforms a baseline model using only audio input the most, where the overall accuracy grows from 77.56\% to 78.04\%. Experimental results also show that using all the subject information, including age, gender, and the specifications of headphones or earphones, is more effective than using only a part of them.
\end{abstract}
\section{Introduction}
\label{sec:intro}

Automatic audio quality measurement can be applied in many different audio related tasks, including detecting fake-quality audio \cite{zhou2014detecting}, monitoring speech quality for online conference \cite{yi2022conferencingspeech}, or assessing the naturalness of synthesized speech \cite{patton2016automos}. Besides, since many people use mobile phones in their daily life, and audio quality is one of the considerations of choosing mobile phones \cite{sriram2019mobile}, measuring the quality of audio playback for mobile phones would also be one of the important tasks.

On the other hand, while some research focused on predicting the mean of users' scores \cite{manocha2022sqapp}, methods that aimed to make personalized predictions have also been studied. Saquil et al. classified users into different categories and used the categorical labels as one of the inputs of their network to rank the preference of different video segments \cite{saquil2021multiple}. Xu et al. added personal information and personality test results of subjects to the input features to their models to predict the preference of sad music \cite{xu2021predicting}.

While Saquil et al. \cite{saquil2021multiple} and Xu et al. \cite{xu2021predicting} focused on the preference of different content, Leng et al. \cite{leng2021mbnet} and Hunag et al. \cite{huang2022ldnet} focused on the preference prediction of the same content in different qualities. Leng et al. proposed MBNet, which was composed of a MeanNet accepting only audio spectrum to predict the mean opinion score (MOS), and a BiasNet in parallel with MeanNet accepting both audio spectrum and subject's ID to predict the possibly biased scores judged by individual subjects \cite{leng2021mbnet}. To improve MBNet, Hunag et al. \cite{huang2022ldnet} proposed LDNet, which used an encoder-decoder structure to predict the listener-dependent score, where the encoder accepted the input speech, and the decoder accepted the encoded vector and the subject's ID to output the listener-dependent score. However, the largest limitation of their methods is that the input of the subject's information is simply their ID, thus the subjects cannot be unseen at the inference stage.

To our best knowledge, despite the impact of subject information on listening tests has been discussed \cite{walton2018role}, there is no research focused on adding subjects' personal information to the prediction model for audio quality preference. In this paper, we aim to predict the personalized preference of two audio segments played by different mobile phones, i.e. preference of two audio segments with the same content in different qualities. Model performance of using several network structures and types of personal information are investigated. The rest of this paper is organized as follows. Section \ref{sec:dataset} describes the process of dataset collection, Section \ref{sec:methods} describes our proposed methods, Section \ref{sec:exp} shows the experimental setting and results, and Section \ref{sec:con} concludes this paper and addresses future work.

\section{Dataset}
\label{sec:dataset}

To obtain audio clips of different qualities, we select some song segments and use some devices to play the segments, and record them. In this study, seven song segments selected from English, Mandarin, Japanese, and Korean pop songs are used for playback by five mobile phones with different brands. Each song segment is about 10 to 15 seconds long and is played by each mobile phone with two different volumes. The first volume is each individual phone’s max volume, which is about 78 to 84 dB for different phones. The second volume is called normal volume, which is about 70 to 75 dB for different phones. A 3DIO Free Space XLR binaural microphone and an ART USB Dual Pre Project Series computer interface are used for recording. Configurations for recording, including the distance between mobile phone and microphone and volume gain, are kept the same for all song segments and all mobile phones.

The recorded files are then used to make questionnaires. The five recorded audio files of each volume of each song segment form ten pairs, where each of the two audio files in one pair corresponds to different mobile phones, and the order of the two audio files is randomly assigned. Therefore, each of the two volumes has 70 pairs. These 70 pairs are divided into five questionnaires, where each questionnaire contains 14 pairs corresponding to the seven different songs. In each pair of each questionnaire, subjects will choose which one is better according to their own preference. The preference score is an integer in [1, 5], where 1 and 2 indicates that the subject prefers the first audio, 3 indicates that no preference for any one of the two audio files, and 4 and 5 indicates that the subject prefers the second audio. For convenience, we will translate the scores to the range of [-2, 2] for further description and analysis.


31 subjects joined the test, including 27 males and 4 females. Their ages range from 21 to 46 years, where 21 of them are between the ages of 21 and 25. The total number of questions answered by subjects is 2,856. Beside, subjects are asked to provide the brand and model of headphones or earphones they used for this test, and the specifications of headphones or earphones, including impedance, lower and upper bound of frequency response, and sensitivity will be collected from the internet. After removing the data of subjects where more than half of the specifications of headphones or earphones they used cannot be found, and the questions of which preference score were answered as 0 (i.e. no preference for any one of the two audio files), the number of remaining subjects is 23, and the number of audio pairs being answered is 2,000. The number of answers of $\pm 2$ and $\pm 1$ are respectively 936 and 1,064.

\section{Methods}
\label{sec:methods}

\begin{figure}[t!]
    \centering
    \includegraphics[width=9cm]{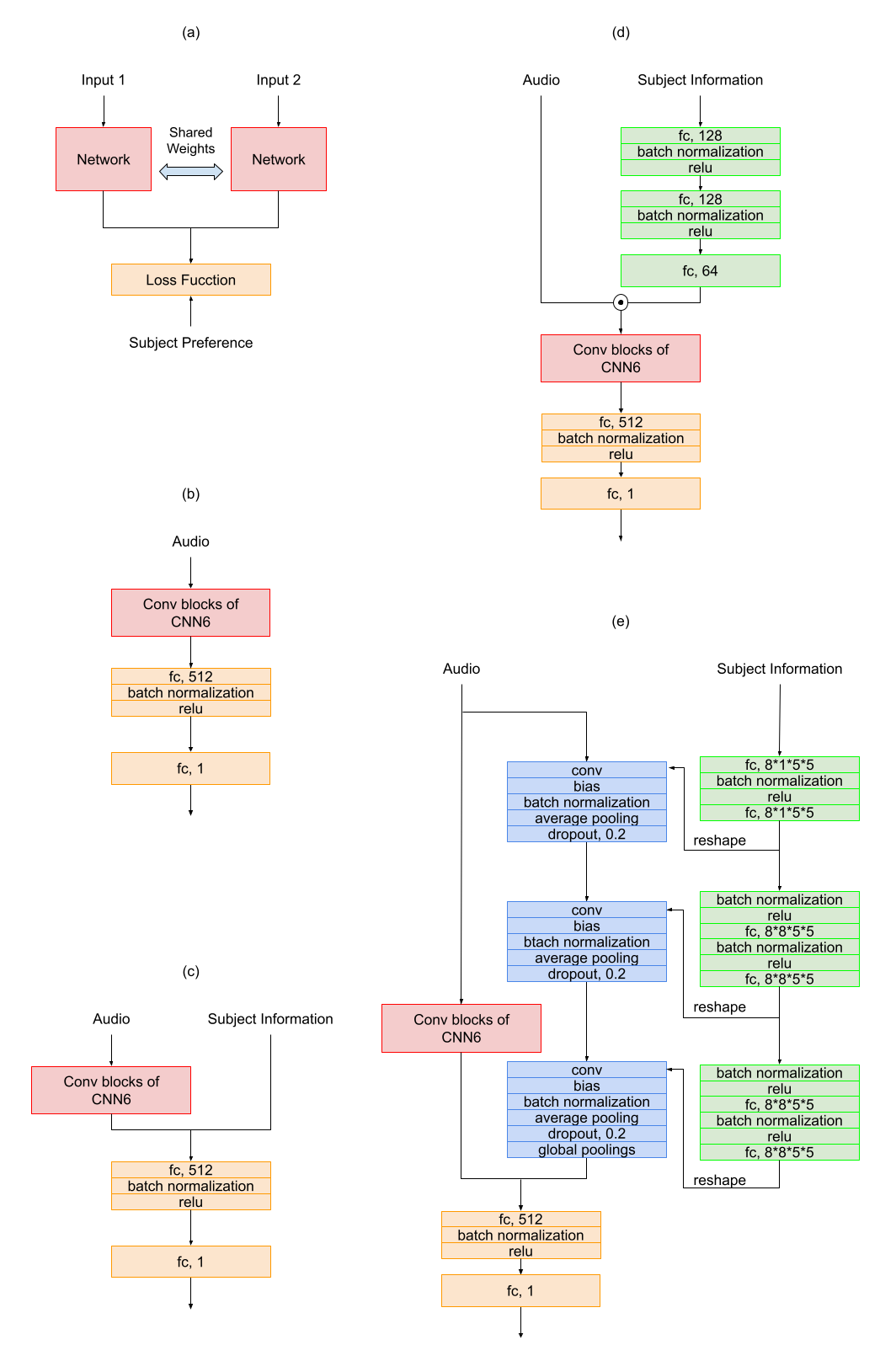}
    \caption{Model structure. (a) Siamese network. (b) The network uses audio input only (\textbf{AO}). (c) The network uses both audio input and subject information, fused later than the CNN6 (\textbf{A+S-L}). (d) Audio input and subject information fused earlier than the CNN6 (\textbf{A+S-E}). (e) Subject information is used as the kernels of convolution to process audio input in parallel with CNN6 (\textbf{A+S-P}).}
    \label{fig:models}
\end{figure}

Since the main task of this work is to predict the preference between two audio segments, the siamese network structure is used to make the output preference the same even when the order of input is changed. Fig. \ref{fig:models}a shows the overall structure of the siamese network, where the inputs are passed through two networks sharing the same weights, and the outputs of the two networks are concatenated together, passed through a softmax function, and used for further loss calculation. For the two sides of the siamese network, the side with the larger output means that the corresponding input audio is preferred by a subject. Note that the network is trained to predict the preference of qualities of two audio segments for one subject, thus the audio segment input to the two sides are in different qualities, but the subject information input to the two sides are the same.

Fig. \ref{fig:models}b shows the network which uses only audio features on each side. The pretrained model ``CNN6'' in PANNs \cite{kong2020panns}, which accepts a log-mel spectrogram of 64 mel bins as input, is used as the basic structure. The original fully-connected layers of CNN6 were replaced with a multi-layer perceptron (MLP) block containing a linear with 512 output nodes, a batch normalization layer, an activation function of ReLU, and a linear layer with one output node. We name this structure \textbf{AO} in this paper.

Fig. \ref{fig:models}c shows the first network using both audio input and subject information on each side. The subject information contains one dimension of age, one dimension of gender, and four dimensions of specifications of headphones or earphones, i.e. impedance, lower and upper bound of frequency response, and sensitivity. The value of $-1$ will be used for missing values of specifications of headphones or earphones. The six-dimensional vector is concatenated with the output of the convolutional block of CNN6, and fed into consequent layers together. This network can be viewed as an LDNet with CNN6 as the encoder and an MLP block as the decoder, except that the input subject information is not only their ID, and this network will be combined with another identical one and viewed as a siamese network. We name this structure \textbf{A+S-L} (L for late fusion) in this paper.

To investigate the possibility of different methods for using both audio input and subject information, two more structures of networks are introduced in this paper. Fig. \ref{fig:models}d and Fig. \ref{fig:models}e respectively show the second and the third network using both audio input and subject information on each side. In the network in Fig. \ref{fig:models}d, the subject information is passed through an MLP block, multiplied with audio input, and then fed into remaining parts. We name this structure \textbf{A+S-E}, where E is for early fusion. In the network in Fig. \ref{fig:models}e, the subject information is also passed through some MLP blocks, but the output of these MLP blocks are reshaped and used as the convolution kernel for input audio, and the convolutional blocks are designed to be similar to that of CNN6. This path is in parallel with the original CNN6, and the outputs of this path and CNN6 will be concatenated together for further processing. We name this structure \textbf{A+S-P}, where P is for parallel.

The sample rate of audio files are originally 44.1 kHz and are reampled to 32 kHz for the use of CNN6. Frame size and hop size are respectively 32 ms and 10 ms. SpecAugment \cite{park2019specaugment} is applied for the training data online during the training process for data augmentation, where the number of stripes for both time and frequency axes is 2, and the maximal drop widths are 64 for the time axis and 4 for the frequency axis. The negative log-likelihood is used as the loss function, and the ADAM optimizer is used to optimize the network. The initial learning rate is $5e-4$ and will be decreased by a factor of 0.95 after each epoch. The batch size is 64. In all of the models, the CNN6 will be fine-tuned.

\section{Experimental Results}
\label{sec:exp}

\subsection{Experimental Setup}
\label{ssec:setup}

To test the ability of predicting the preference for unseen subjects, we divide all the 23 subjects into 7 folds. All the subjects are sorted by their age, and when the subjects of index $k$ are used for testing, the subjects of index $k+1\,\,mod\,\,7$ are used for validation, and the remaining subjects are used for training. The number of epochs for training is 50, and the training process will be early-stopped if the validation loss does not achieve a new lower value for consequently 10 epochs. The evaluation metric is accuracy, which is defined by the number of pairs of which the preferences are correctly predicted divided by the total number of test pairs.

Experiments are conducted in a container running on a machine with Intel(R) Xeon(R) Gold 6154 CPU, occupying 4 CPU cores, 90 GB of host  memory, and a Tesla V100-SXM2 GPU.

\subsection{Results}
\label{ssec:results}

We first compare the performances of different model. The test accuracies of the 7 folds and the overall accuracies are shown in Table \ref{tab:results_1}, where the results of each model in each fold are the average of 21 runs, and the overall accuracies are calculated by the weighted average of accuracies of all the folds. As shown in the table, \textbf{A+S-L} performs better than \textbf{AO} on average, where the accuracy grows from 77.56\% to 78.04\%. For individual folds, higher mean values are achieved in 5 out of the 7 folds, lower standard deviations are achieved in 6 out of the 7 folds, and mean values are higher and standard deviations are lower in 4 out of the 7 folds. On the other hand, performance of \textbf{A+S-E} is significantly lower than other models, and \textbf{A+S-P} achieves similar mean values to \textbf{AO} and \textbf{A+S-L}, but the standard deviations are higher, which is possibly due to that the network is more complex but the amount of data is not very large. We do not compare to other LDNet structures which use different encoders and decoders because the amount of data is currently not very large, thus may not be suitable for training more complex network structures.

\begin{table*}[t]
  \centering
  \begin{tabular}{ccccc}
    \hline
    Fold (\#Q) \textbackslash~Model & AO & A+S-L & A+S-E & A+S-P \\
    \hline
    \hline
    1 (403) & 0.8029 $\pm$ 0.0151 & \textbf{0.8056 $\pm$ 0.0069} & 0.5757 $\pm$ 0.0462 & 0.8098 $\pm$ 0.0188 \\
    2 (269) & 0.7836 $\pm$ 0.0121 & \textbf{0.7946 $\pm$ 0.0062} & 0.6887 $\pm$ 0.1071 & 0.7874 $\pm$ 0.0132 \\
    3 (318) & \textbf{0.7843 $\pm$ 0.0118} & 0.7808 $\pm$ 0.0066 & 0.6499 $\pm$ 0.1241 & 0.7719 $\pm$ 0.0117 \\
    4 (215) & \textbf{0.6742 $\pm$ 0.0134} & 0.6670 $\pm$ 0.0118 & 0.5826 $\pm$ 0.0698 & 0.6684 $\pm$ 0.0151 \\
    5 (289) & 0.7569 $\pm$ 0.0157 & \textbf{0.7670 $\pm$ 0.0091} & 0.6753 $\pm$ 0.0537 & 0.7161 $\pm$ 0.0539 \\
    6 (305) & 0.7761 $\pm$ 0.0063 & \textbf{0.7864 $\pm$ 0.0076} & 0.5633 $\pm$ 0.1340 & 0.7775 $\pm$ 0.0069 \\
    7 (201) & 0.8316 $\pm$ 0.0171 & \textbf{0.8420 $\pm$ 0.0141}  & 0.5714 $\pm$ 0.0538 & 0.8380 $\pm$ 0.0164  \\
    \hline
    Overall & 0.7756 & \textbf{0.7804} & 0.6155 & 0.7699 \\
    \hline
  \end{tabular}
  \caption{The test accuracies of the 7 folds and the overall accuracies for different models. Number of preference questions answered by subjects is denoted by ``\#Q''. The results of each model in each fold are the average of 21 runs. The mean value and standard deviation is denoted by the form of ``mean$\pm$std''.}
  \label{tab:results_1}
\end{table*}

\begin{figure}[t!]
    \centering
    \includegraphics[width=8.5cm]{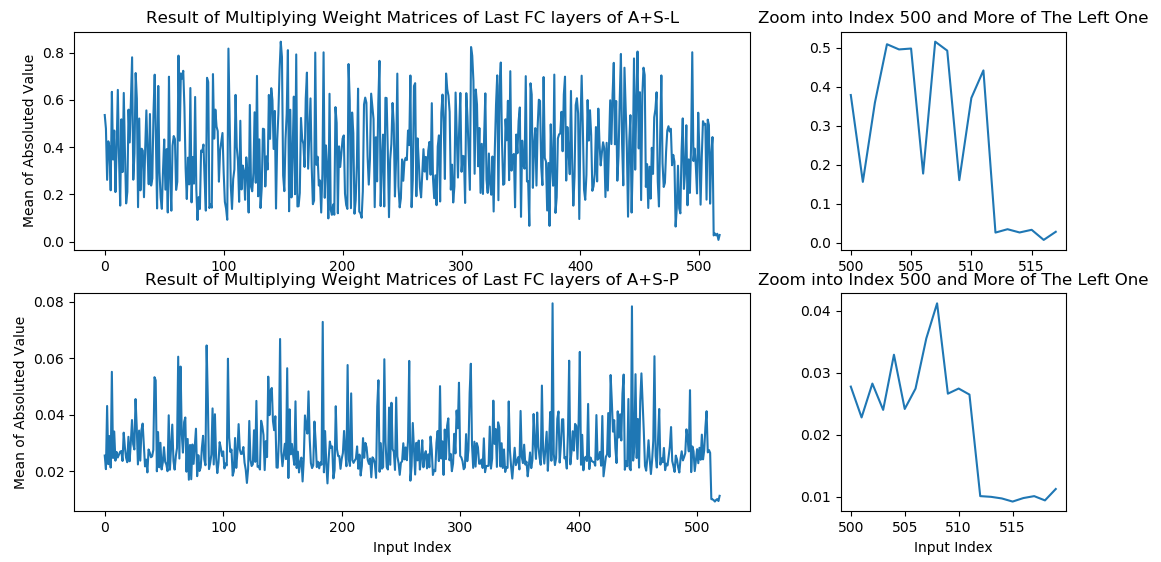}
    \caption{Mean absolute values of multiplied results of weight matrices of the linear layers of the last MLP block of \textbf{A+S-L} (upper plots) and \textbf{A+S-P} (lower plots), where the left and right plots are respectively all the values and values of and after index 500.}
    \label{fig:mwm}
\end{figure}

To observe the importance of both audio input and subject information, the weight matrices  of the linear layers of the last MLP block of \textbf{A+S-L} and \textbf{A+S-P} are respectively multiplied, and the mean absolute values are shown in Fig. \ref{fig:mwm}, where the first 512 values are for audio information processed by CNN6, and the remaining 6 or 8 values are for raw subject information in \textbf{A+S-L} or subject information convolved with audio information in \textbf{A+S-P}. For \textbf{A+S-L}, the first 512 values are around 0.4, and the remaining values are around 0.02. This difference of scales may simply be due to the fact that the values of audio information processed by CNN6 are normalized but the values of raw subject information are not. Therefore the importance of audio and subject information may be similar. For \textbf{A+S-P}, both audio and subject information are normalized, but the mean absolute values of multiplied weight matrices are at different levels (0.03 for audio and 0.01 for subject information), so the subject information may not be utilized very well in \textbf{A+S-P}, but collecting more data to train a more complex model may improve the accuracy.

\begin{table*}[t]
  \centering
  \begin{tabular}{cccccc}
    \hline
    Fold (\#Q) \textbackslash~\makecell{Subject \\ Info} & all & age \& gender & \makecell{all specs of \\ head/earphones} & \makecell{impd \& sensit \\ of head/earphones} & \makecell{freq responses \\ of head/earphones} \\
    \hline
    \hline
    1 (403) & 0.8056 $\pm$ 0.0069 & 0.8117 $\pm$ 0.0177 & 0.8060 $\pm$ 0.0103 & \textbf{0.8134 $\pm$ 0.0197} & 0.8073 $\pm$ 0.0154 \\
    2 (269) & 0.7946 $\pm$ 0.0062 & 0.7775 $\pm$ 0.0134 & \textbf{0.7959 $\pm$ 0.0098} & 0.7809 $\pm$ 0.0104 & 0.7864 $\pm$ 0.0059 \\
    3 (318) & 0.7808 $\pm$ 0.0066 & 0.7811 $\pm$ 0.0108 & 0.7767 $\pm$ 0.0089 & \textbf{0.7848 $\pm$ 0.0224} & 0.7774 $\pm$ 0.0081 \\
    4 (215) & 0.6670 $\pm$ 0.0118 & \textbf{0.6712 $\pm$ 0.0107} & 0.6709 $\pm$ 0.0114 & 0.6668 $\pm$ 0.0183 & 0.6633 $\pm$ 0.0135 \\
    5 (289) & 0.7670 $\pm$ 0.0091 & 0.7330 $\pm$ 0.0113 & \textbf{0.7690 $\pm$ 0.0108} & 0.7631 $\pm$ 0.0255 & 0.7681 $\pm$ 0.0093 \\
    6 (305) & 0.7864 $\pm$ 0.0076 & 0.7807 $\pm$ 0.0071 & \textbf{0.7866 $\pm$ 0.0064} & 0.7672 $\pm$ 0.0121 & 0.7831 $\pm$ 0.0070 \\
    7 (201) & \textbf{0.8420 $\pm$ 0.0141}  & 0.8224 $\pm$ 0.0138 & 0.8398 $\pm$ 0.0094 & 0.8403 $\pm$ 0.0181 & 0.8400 $\pm$ 0.0138 \\
    \hline
    Overall & 0.7804 & 0.7721 & \textbf{0.7806} & 0.7771 & 0.7782  \\
    \hline
  \end{tabular}
  \caption{Test accuracies of \textbf{A+S-L} using different subsets of subject information. \textbf{impd}: impedance. \textbf{sensit}: sensitivity. \textbf{freq}: frequency.}
  \label{tab:results_2}
\end{table*}


Test accuracies of \textbf{A+S-L} using different subsets of subject information are shown in Table \ref{tab:results_2}. When using only age and gender as the subject information, the model performs slightly worse than that of using other subsets of subject information in most of the folds. On the other hand, when using all specifications of headphones or earphones as the subject information, the accuracies are slightly higher than that of using all the available subject information, but the standard deviations are also slightly higher. Above results suggest that the model performs more stable when all the available subject information is used.


\section{Conclusions and future work}
\label{sec:con}

This paper proposes to make personalized prediction of the preference of two audio segments with the same content but are in different qualities using both audio input and subject information. Siamese network is used to compare two sets of input, and the side with larger output indicates that the corresponding input audio is preferred. For each side of the siamese network, the performances of several different structures are investigated, and an LDNet \cite{huang2022ldnet} with CNN6 \cite{kong2020panns} as the encoder and an MLP block as the decoder outperforms a baseline model using only audio input the most, where the overall accuracy grows from 77.56\% to 78.04\%. The effectiveness of different types of subject information is also investigated, and using all the available subject information, including age, gender, and the specifications of headphones or earphones, is found to be the most stable. Since currently the ages of most subjects concentrate in a small range, recruiting more subjects with different ages in the future may help with a better discussion of the effectiveness of age to predict the personalized preference of audio quality. Besides, since the environment and other information for listening devices including life and power of headphones or earphones may also affect the audio preference, collecting such information may also help predict the audio preference better.

\section{Acknowledgements}
\label{sec:ack}

This work was supported by the ASUSTek Computer Inc.. We also thank the National Center for High-performance Computing (NCHC) for providing computational and storage resources.



\bibliographystyle{IEEEbib}
\balance
\bibliography{Template}

\begin{thebibliography}{10}

\bibitem{zhou2014detecting}
Jinglei Zhou, Rangding Wang, Chao Jin, and Diqun Yan,
\newblock ``Detecting fake-quality wav audio based on phase differences,''
\newblock in {\em International Workshop on Digital Watermarking}. Springer,
  2014, pp. 525--534.

\bibitem{yi2022conferencingspeech}
Gaoxiong Yi, Wei Xiao, Yiming Xiao, Babak Naderi, Sebastian M{\"o}ller, Wafaa
  Wardah, Gabriel Mittag, Ross Cutler, Zhuohuang Zhang, Donald~S Williamson,
  et~al.,
\newblock ``Conferencingspeech 2022 challenge: Non-intrusive objective speech
  quality assessment (nisqa) challenge for online conferencing applications,''
\newblock {\em arXiv preprint arXiv:2203.16032}, 2022.

\bibitem{patton2016automos}
Brian Patton, Yannis Agiomyrgiannakis, Michael Terry, Kevin Wilson, Rif~A
  Saurous, and D~Sculley,
\newblock ``Automos: Learning a non-intrusive assessor of
  naturalness-of-speech,''
\newblock {\em arXiv preprint arXiv:1611.09207}, 2016.

\bibitem{sriram2019mobile}
KV~Sriram, H~Mahesh Prabhu, and Aditi~Ajay Bhat,
\newblock ``Mobile phone usability and its influence on brand loyalty and
  re-purchase intention: An empirical,''
\newblock in {\em 2019 IEEE International WIE Conference on Electrical and
  Computer Engineering (WIECON-ECE)}. IEEE, 2019, pp. 1--4.

\bibitem{manocha2022sqapp}
Pranay Manocha, Zeyu Jin, and Adam Finkelstein,
\newblock ``Sqapp: No-reference speech quality assessment via pairwise
  preference,''
\newblock in {\em ICASSP 2022-2022 IEEE International Conference on Acoustics,
  Speech and Signal Processing (ICASSP)}. IEEE, 2022, pp. 891--895.

\bibitem{saquil2021multiple}
Yassir Saquil, Da~Chen, Yuan He, Chuan Li, and Yong-Liang Yang,
\newblock ``Multiple pairwise ranking networks for personalized video
  summarization,''
\newblock in {\em Proceedings of the IEEE/CVF International Conference on
  Computer Vision}, 2021, pp. 1718--1727.

\bibitem{xu2021predicting}
Liuchang Xu, Ye~Zheng, Dayu Xu, and Liang Xu,
\newblock ``Predicting the preference for sad music: the role of gender,
  personality, and audio features,''
\newblock {\em IEEE Access}, vol. 9, pp. 92952--92963, 2021.

\bibitem{leng2021mbnet}
Yichong Leng, Xu~Tan, Sheng Zhao, Frank Soong, Xiang-Yang Li, and Tao Qin,
\newblock ``Mbnet: Mos prediction for synthesized speech with mean-bias
  network,''
\newblock in {\em ICASSP 2021-2021 IEEE International Conference on Acoustics,
  Speech and Signal Processing (ICASSP)}. IEEE, 2021, pp. 391--395.

\bibitem{huang2022ldnet}
Wen-Chin Huang, Erica Cooper, Junichi Yamagishi, and Tomoki Toda,
\newblock ``Ldnet: Unified listener dependent modeling in mos prediction for
  synthetic speech,''
\newblock in {\em ICASSP 2022-2022 IEEE International Conference on Acoustics,
  Speech and Signal Processing (ICASSP)}. IEEE, 2022, pp. 896--900.

\bibitem{walton2018role}
Tim Walton and Michael Evans,
\newblock ``The role of human influence factors on overall listening
  experience,''
\newblock {\em Quality and User Experience}, vol. 3, no. 1, pp. 1--16, 2018.

\bibitem{kong2020panns}
Qiuqiang Kong, Yin Cao, Turab Iqbal, Yuxuan Wang, Wenwu Wang, and Mark~D
  Plumbley,
\newblock ``Panns: Large-scale pretrained audio neural networks for audio
  pattern recognition,''
\newblock {\em IEEE/ACM Transactions on Audio, Speech, and Language
  Processing}, vol. 28, pp. 2880--2894, 2020.

\bibitem{park2019specaugment}
Daniel~S Park, William Chan, Yu~Zhang, Chung-Cheng Chiu, Barret Zoph, Ekin~D
  Cubuk, and Quoc~V Le,
\newblock ``Specaugment: A simple data augmentation method for automatic speech
  recognition,''
\newblock {\em arXiv preprint arXiv:1904.08779}, 2019.

\end{thebibliography}

\end{document}